\documentclass[11pt,a4paper]{article}
\pdfoutput=1

\usepackage{jcappub}

\usepackage{amsmath}
\usepackage[normalem]{ulem}
\usepackage{graphicx}
\DeclareGraphicsExtensions{.eps, .eps, .jpg, .png}

\def\bk{{\bf k}}

\def\bp{{\bf p}}
\def\la{~\mbox{\raisebox{-.6ex}{$\stackrel{<}{\sim}$}}~}

\title{Suppressing the impact of a high tensor-to-scalar ratio on the temperature anisotropies}

\author[a,b]{Carlo R. Contaldi}

\author[c]{Marco Peloso}

\author[d,e]{Lorenzo Sorbo}

\affiliation[a]{Theoretical Physics, Blackett Laboratory, Imperial College, London, SW7 2BZ, UK}
\affiliation[b]{Canadian Institute of Theoretical Physics, Toronto, M5S 3H8, On, Canada}
\affiliation[c]{School of Physics and Astronomy, University of Minnesota, Minneapolis, 55455, USA}
\affiliation[d]{Department of Physics, University of Massachusetts, Amherst, MA 01003, USA}
\affiliation[e]{Institut de Physique Th\'eorique, CEA, F-91191 Gif-sur-Yvette Cedex, France}

\emailAdd{c.contaldi@imperial.ac.uk, peloso@physics.umn.edu, sorbo@physics.umass.edu}

\abstract{
The BICEP2 collaboration has reported a strong $B$ mode signal in
  the CMB polarization, which is well fit by a tensor-to-scalar ratio
  of $r\simeq 0.2$. This is greater than the upper limit $r < 0.11$
  obtained from the temperature anisotropies under the assumption of a
  constant scalar spectral index $n_s$. This discrepancy can be
  reduced once the statistical error and the contamination from
  polarized dust are accounted for. If however a large value for $r$
  will be confirmed, it will need to be reconciled  with the
  temperature anisotropies data.  The most advocated explanation
  involves a variation of $n_s$  with scales  (denoted as running) that has a magnitude
  significantly greater than the generic slow roll predictions. We
  instead study the possibility that the large scale temperature
  anisotropies are not enhanced because of a suppression of the
  scalar power at large scales. Such a situation can be achieved for
  instance by a sudden change of the speed of the inflaton (by about $14\%$), and we show that
  it fits the  temperature anisotropies and polarization data considerably better than a constant 
  running (its $\chi^2$ improves by $\sim 7.5$ over that of the constant running, at the cost of
  one more parameter). We also consider the possibility that the large scale 
  temperature fluctuations are suppressed by an anti-correlation between
  tensor and scalar modes. Unfortunately, while such effect does affect the temperature fluctuations
  at large scales, it does not affect the temperature power spectrum and cannot, 
  therefore, help in reconciling a large value of $r$ with the limits from temperature fluctuations.
   }

\begin{document}
\maketitle

\section{Introduction}

The BICEP2 experiment \cite{Ade:2014xna} has observed a $B$-mode polarization of the
Cosmic Microwave Background (CMB)
that can be well fit by a lensed-$\Lambda$CDM + tensor theoretical
model, with tensor-to-scalar ratio $r=0.2^{+0.07}_{-0.05}$, with $r=0$
disfavored at $7\sigma$.  It is possible that the actual primordial
component of the $B$-mode found by BICEP2 is smaller than $r=0.2$.  A
number of tests were performed on the BICEP2 data to ensure that the
the observed value is not due to any instrumental effects. Moreover, the
lensing contribution to $B$-modes does not appear to be sufficiently
large to explain the measured value. The signal, observed by BICEP2, peaks at $\ell \approx 100$, where the
primordial signal is expected to dominate, whilst the lensing signal peaks at $\ell
\approx 1000$. 

Other potential contaminants are the Galactic synchrotron and
polarized-dust emission. Whilst the former effect is negligible at the
BICEP2 observing frequency, the polarized dust is a substantial
contaminant.  Although the  area of sky probed by BICEP2 is very clean with
respect to the total intensity emission by dust there is still  much uncertainty
in the level of polarised dust contamination due to the lack of
observations. A number of models were considered by BICEP2 for the
subtraction of the dust contamination which result in some shift in
the maximum likelihood values in $r$. A further argument 
for the primordial nature of the signal is that cross-correlation
between frequencies (specifically, the preliminary BICEP2 $\times$ Keck
cross-correlation shown in  \cite{Ade:2014xna}) displays little change in the observed
amplitude. This appears to indicate  that frequency dependent foregrounds
are not the dominant contributor to the observed $B$-modes. 
On the other hand, more recent works \cite{antibicep} (appeared after the first version of this manuscript) argue that the polarized dust contamination is likely stronger than what assumed in  \cite{Ade:2014xna}. 
This could invalidate BICEP2 claim of a primordial origin of the observed signal, or at least reconcile it with the Planck limit \cite{Ade:2013uln} $r < 0.11$. In this work we assume that the BICEP2 noise estimate is correct. The data
already collected by Planck should have the sensitivity and frequency
range to definitely confirm or rule out the primordial nature of such a large signal.

If the primordial contribution to the $B$-modes is confirmed, it will 
represent the first detection of gravity waves from inflation
\cite{Ade:2014xna}. This is of paramount importance, since - under the
assumption that the observed gravity waves are those created by a
period of 
quasi de-Sitter inflationary expansion (namely that $P_t \sim V/M_p^4$, where $P_t$ is the tensor power spectrum,
$V^{1/4}$ is the energy scale of inflation, and $M_p \simeq 2.4 \times
10^{18} $ GeV is the reduced Planck mass) - it allows us, for the first
time, to determine the energy scale of inflation. From the
parametrization $r \equiv P_t/P_s$, and from the measured
value of the scalar power spectrum, $P_s \simeq 2.45 \times 10^{-9}$,
one obtains the well known relation
\begin{equation}
V^{1/4} \simeq 2.25 \cdot 10^{16} \, {\rm GeV} \; \left( \frac{r}{0.2} \right)^{1/4} \;. 
\end{equation}
Therefore, if the $B$-mode signal observed by BICEP2 is due to
inflationary vacuum modes, we have now learnt that inflation took
place at the GUT scale.

As we already mentioned, taken at face value, the BICEP2 value is in strong tension with the $2
\sigma$ limit $r<0.11$ obtained by the Planck inflation analysis
\cite{Ade:2013uln}. Such a limit however relies on the scaling of the
temperature anisotropy data (supplemented by the WMAP large-scale
polarization likelihood), and not on the direct measurement of the
$B$-mode polarization.  The $r<0.11$ limit appears robust under the
inclusion of several data sets (such as the ACT+SPT temperature data, BAO,
and the Planck lensing \cite{Ade:2013uln}). However, it crucially
relies in the assumption of a constant spectral tilt $n_s$.

Specifically, it is obtained from the Planck+ACT+SPT temperature data
(with the Planck data supplemented by the WMAP large-scale
polarization likelihood), under the assumption of constant spectral
tilt $n_s = 0.960 \pm 0.007$ \cite{Ade:2013uln}. As discussed in
\cite{Ade:2013uln}, a more relaxed limit is obtained if $n_s$ is
allowed to vary with scale $k$. Specifically, it is customary to parametrize the
scalar power spectrum as
\begin{equation}
P_s \left( k \right) \equiv P \left( k_0 \right) \left( \frac{k}{k_0} \right)^{n_s - 1 + \frac{1}{2} \alpha_s \ln \frac{k}{k_0} } \;,
\label{parametrizationP}
\end{equation}
where $k_0 = 0.05$ Mpc$^{-1}$, is the chosen pivot scale (this
is the scale at which also $r$ is defined) and the parameter
$\alpha_s$ denotes the running of the scalar spectral tilt
\cite{Kosowsky:1995aa} with $\alpha_s = \frac{d \, n_s}{d \, {\rm ln }
  \, k}$.

If $\alpha_s \neq 0$, the $r<0.11$ limit is relaxed to $r \la
0.25$. From Figure~5 of \cite{Ade:2013uln} we infer that a value
$\alpha_s \sim - 0.02$ is required to reconcile the temperature data
with $r =0.2$. Such a large value of $\vert \alpha_s \vert$ is not a
generic prediction of slow roll inflationary models. Indeed, in terms
of the slow roll parameters
\begin{equation}
\epsilon \equiv \frac{M_p^2}{2} \left( \frac{V_{,\phi}}{V} \right)^2 \;,\; 
\eta \equiv M_p^2 \frac{V_{,\phi\phi}}{V} \;,\;
\xi^2 \equiv M_p^4 \frac{V_{,\phi} V_{,\phi\phi\phi}}{V^2} \;,
\end{equation} 
where $V$ denotes the potential of the inflaton $\phi$ and comma denotes a derivative, we have the well known slow roll relations 
\begin{eqnarray}
&& r = 16 \, \epsilon \;\;\;,\;\;\; n_s - 1 = 2 \eta - 6 \epsilon\,, \nonumber\\ 
&& \alpha_s = - 2 \xi^2 + \frac{r}{2} \left( n_s - 1 \right) +
\frac{3}{32} r^2 \simeq - 2 \xi^2 - 0.00025\,, \nonumber\\ 
\label{slow-roll}
\end{eqnarray}
where $n_s = 0.96$, $r=0.2$ has been used in the final numerical
estimate. This is typically much smaller than the required value,
since, as evident in (\ref{slow-roll}), the running is generically of
second order in slow roll.

In principle, models can be constructed in which the third derivative
term $\xi^2$ is ``anomalously large''. However, besides being hard to
motivate, it is difficult to maintain a large third derivative, while
the first two derivatives are small, for a sufficiently long duration
of inflation, \cite{Chung:2003iu,Easther:2006tv}, so that the models
in which a large running is achieved have potentials with some
bump-like feature or superimposed oscillations
\cite{Hannestad:2000tj,Chung:2003iu,
  Feng:2003mk,Kobayashi:2010pz,Takahashi:2013tj}, or possess some
peculiar aspects beyond standard scenarios
\cite{Kawasaki:2003zv,Huang:2003zp,BasteroGil:2003bv,Yamaguchi:2003fp,Ballesteros:2005eg}.

In summary, it appears that $r=0.2$ can be reconciled with the limits
from the temperature anisotropies through a negative running, which is
however of substantially larger magnitude than the generic slow roll
prediction. It is possible that the value of $r$ from the polarization
will shift towards $r \sim 0.1$, in which case the tension with the
temperature data can be relaxed (or disappear altogether). This can
happen factoring in both the statistical uncertainty in the BICEP2
$r=0.2^{+0.07}_{-0.05}$ result, and the decrease of $r$ that appears
in most of the model-dependent dust corrections \cite{Ade:2014xna}.

Remarkably, $r$ close to $0.15$ appears as a prediction of the
simplest models of inflation, such as chaotic
inflation~\cite{Linde:1983gd} and natural
inflation~\cite{Freese:1990rb}. Even if UV complete theories typically
leads to a lower inflationary scale, it is possible to construct
models that can evade the fundamental constraints which typically make
high-scale inflation difficult to
realize~\cite{Kim:2004rp,Kaloper:2008fb,Kaloper:2011jz,Pajer:2013fsa}
and still display such simple potentials.

However, from a theoretical point of view, it is interesting to
understand the implications that a large measured value for $r$ from
polarization would have for inflationary model building. In this work
we discuss two additional possibilities (in addition to the already mentioned running of the
spectral tilt)  to suppress the large scale temperature signal in presence of a large $r\simeq 0.2$.

The first mechanism relies on the presence of a large scale
suppression in the scalar power. A similar idea was already explored in
\cite{Contaldi:2003zv}, in order to address the suppressed power of
the temperature anisotropies at the largest scales. The best fit to
the first year WMAP data was obtained if the power drops to zero at
scales $k \la 5 \times 10^{-4} \, {\rm Mpc}^{-1}$
\cite{Bennett:2003bz}.  Such a strong suppression can for example
occur if the universe is closed, with a curvature radius comparable to
the horizon at the onset of inflation \cite{Linde:2003hc}, or if the
inflaton was in fast roll at the beginning of the last $\sim 60$
$e$-folds of inflation~\cite{Contaldi:2003zv}. In this case, one also
expects a suppression of the tensor signal at large scales, although
this suppression is milder than that of the scalar power
\cite{Nicholson:2007by}. Here, for simplicity, we only consider a
simple model for suppression in order to explore the viability of such
a model in explaining the surprisingly discrepancy between $TT$ and
$BB$ spectra. We do not assume that the power drastically drops to $\simeq 0$ 
at large scales, but that it decreases by a factor $\left( 1 - \Delta \right)^2$  for
$k$ smaller than a given scale $k_*$ (a similar analysis on Planck data only was performed in~\cite{Hazra:2013nca}). Our best fit is characterized by a $\sim 26\%$ drop in power that can be achieved for example by a change of slope of the inflaton potential, such that the inflaton goes slightly faster when the largest modes were generated. However, the system never leaves the slow-roll regime. Therefore, contrary to the situation studied in \cite{Contaldi:2003zv,Bennett:2003bz} there is no discontinuity in the tensor power. Specifically, we study the model originally proposed by Starobinsky  \cite{Starobinsky:1992ts} for this change in slope (see also~\cite{Kaloper:2003nv}). Alternatively, a change in power can result from    a change of the sound speed of the inflaton perturbations~\cite{Park:2012rh} (see also~\cite{Wu:2006xp,D'Amico:2013iaa} for other models leading to a similar effect).

The second possibility that we discuss is a negative correlation
between the scalar and tensor signal\footnote{In a previous version 
of this manuscript we reached a different conclusion. 
This section has been modified after the findings 
of~\cite{Zibin:2014iea,Emami:2014xga}, see also~\cite{Chen:2014eua}.}. 
This correlation will affect the sum of the scalar 
and tensor modes on the temperature anisotropy. 
Some degree of non-vanishing scalar-tensor correlation is 
expected in the presence of the breaking of Lorentz invariance, and for
example it arises if the background expansion is anisotropic
\cite{Gumrukcuoglu:2006xj,Pereira:2007yy,Gumrukcuoglu:2007bx,Pitrou:2008gk}.
Unfortunately, we will see that such a correlation does not affect 
the temperature power spectrum, and will therefore not help reconcile the constraints on $r$ from $TT$ with those from $BB$ spectra.

The plan of the paper is the following: In Section~\ref{cut-off} we
discuss the effects of a large scale suppression of the scalar
power. In Section~\ref{anti} we discuss the effects of an
anti-correlation between scalar and tensor modes. In Section ~\ref{conclusions}
we present our conclusions. Finally, in Appendix  ~\ref{app} we study a parametrization of the scalar power spectrum characterized by a step function suppression. This parametrization lacks the ``ringing'' effect that is typically encountered in concrete models (see Figure \ref{fig:starob}). The comparison between the two analyses shows that the ringing present in the Starobinsky model has a minor impact on the data.

\section{Suppression of large scale scalar modes}
\label{cut-off}

\begin{figure}[t]
\begin{center}
\includegraphics[width=8.5cm,trim=0cm 0cm 0cm
  0cm,clip]{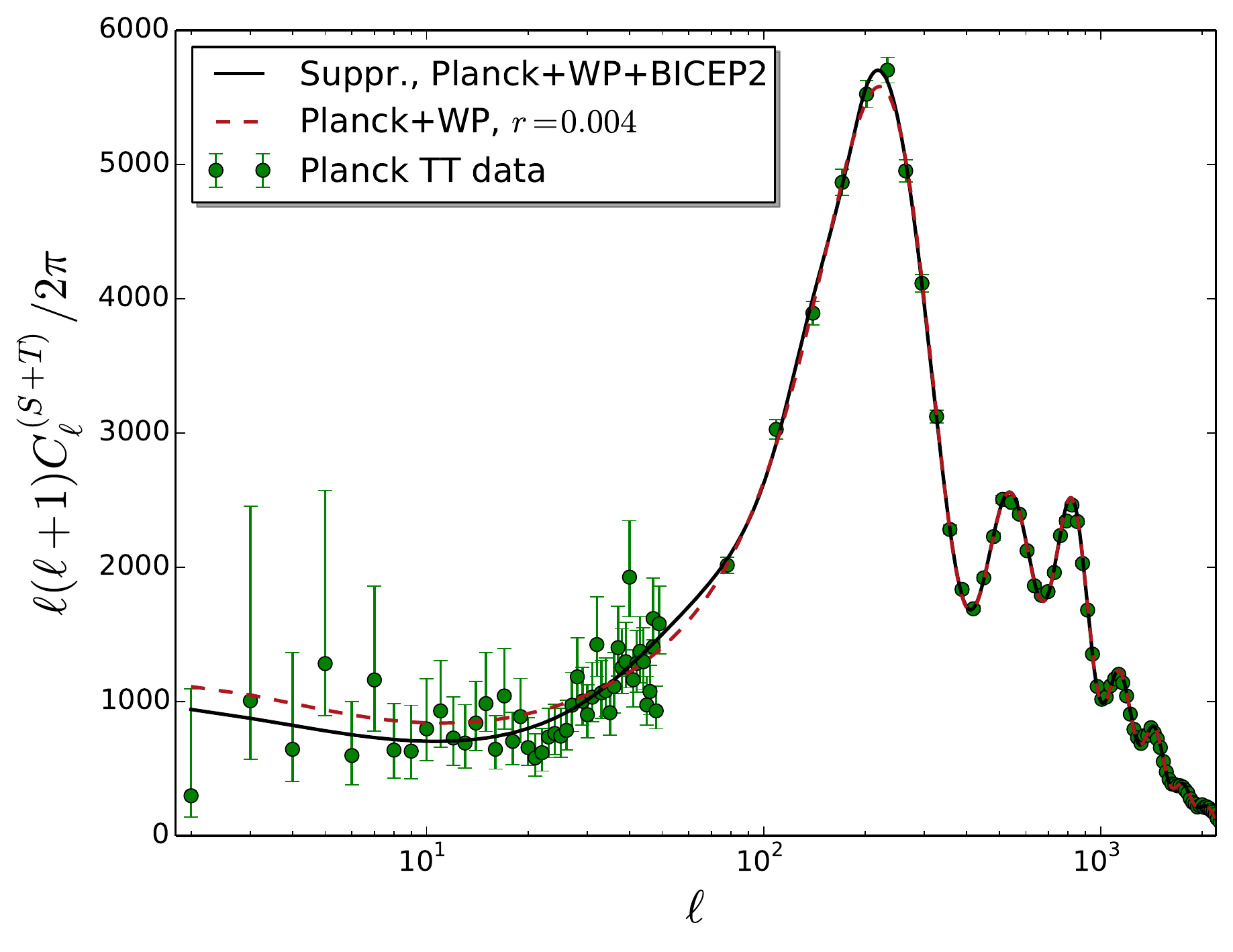}
  \caption{
   The total contribution to the $TT$ spectrum from
  scalar and tensors. The black (solid) curve shows the best-fit
  suppression (or Starobinsky) 
  model with $r=0.18$ from the MCMC chains. The tensor contribution compensates for the loss of
  large-scale scalar power but also allows a better fit to the low
  points in the range $20<\ell<30$. The red (dashed) model is the best-fit
  model from the ``base r'' Planck+WP MCMC runs with
  $r=0.004$. } 
  \label{fig:cltot} 
   \end{center}
\end{figure}

A simple mechanism that produces a suppression at large scales is the Starobinsky model \cite{Starobinsky:1992ts} 
\begin{equation}
V \left( \phi \right) = \left\{ \begin{array}{l} 
V_0 + C_+ \left( \phi - \phi_* \right) \;\;\;,\;\;\; \phi \geq \phi_* \\ 
V_0 + C_- \left( \phi - \phi_* \right) \;\;\;,\;\;\; \phi \leq \phi_* 
\end{array} \right. \;, 
\label{starob}
\end{equation}
where we assume that all the parameters are positive, and $C_- < C_+$. As the inflaton $\phi$ rolls past $\phi_*$ during inflation, it goes from a region of greater ($C_+$) to a region of smaller ($C_-$) slope in the potential, and the scalar perturbations produced in this second stage, namely those with momentum $k > k_*$, where $k_*$ is the momentum of the mode that leaves the horizon when $\phi = \phi_*$, have a greater power than those produced in the first stage, namely, those with momentum $k < k_*$. Defining $\Delta \equiv \frac{C_+-C_-}{C_+}$
and $\kappa\equiv k/k_\star$, one obtains  \cite{Starobinsky:1992ts,Martin:2011sn}:  
\begin{eqnarray}
P_s (k) &=&   A\, \left(\frac{k}{k_0}\right)^{n_s-1} \times \left\{ 1+
    \frac{9\Delta^2}{2}\left(\frac{1}{\kappa} + \frac{1}{\kappa^3}\right)^2 + \frac{3\Delta}{2}\left(4+3\Delta-3\Delta\frac{1}{\kappa^4}\right)\frac{1}{\kappa^2}\cos(2\kappa) \right. \nonumber\\ 
   && \quad\quad\quad\quad\quad\quad +\left. 3\Delta\left[1-(1+3\Delta)\frac{1}{\kappa^2}-3\Delta\frac{1}{\kappa^4}\right]\frac{1}{\kappa}\sin(2\kappa)
\right\}\;, 
\label{P-starob}
\end{eqnarray}
with $A\equiv P(k_0)$, and $k_0$ is a pivot scale. The curly parenthesis in eq. (\ref{P-starob})  contains the precise suppression of power at large scales
\begin{equation}
\lim_{\kappa\to 0} P_s(k) = (1-\Delta)^2\, A\, \left(\frac{k}{k_0}\right)^{n_s-1} \;\;,\;\; 
\lim_{\kappa\to \infty} P_s(k) = A\, \left(\frac{k}{k_0}\right)^{n_s-1} \;, 
\end{equation} 
together with a ``ringing'' in the power that is typically obtained in models that give suppression of the power \cite{Contaldi:2003zv}. We show the power spectrum (\ref{P-starob}) for the best fit to the Planck and BICEP2 data in  Figure \ref{fig:starob}. The best fit value  $\Delta = 0.14$ (see below) results in a $\sim 26\%$ suppression of the power at large wrt short scales. 

We assume that eq. (\ref{starob}) reproduces the potential close to the transition ($\phi \simeq \phi_*$), but we do not necessarily assume that the potential remains linear at all $\phi$. For this reason, in eq. (\ref{P-starob}) we allow for an arbitrary spectral tilt $n_s$. In the data fitting, $n_s$ has an impact on scales $k \gg k_*$, and there is no reason to assume that eq. (\ref{starob}) applies at $\phi$ arbitrarily  far away from $\phi_*$ (in fact, we know that $V$ needs to change from  (\ref{starob}) so to have a minimum with  $V_{\rm min} = 0$). 

Moreover, we stress that the suppression of the power that is considered here is much smaller than the one studied in  \cite{Contaldi:2003zv}. The goal of that work was to essentially suppress all the power at large scales, so to address the problem of the anomalously small observed  quadrupole ($\ell =2$). To achieve this, ref. 
 \cite{Contaldi:2003zv} advocated a period of kinetion domination
 ($\dot{\phi}^2 \gg V$) before the onset of inflation. This also
 resulted in a very different evolution of $H$ in this initial stage,
 and to a modification of the gravity wave power
 \cite{Nicholson:2007by}. In the present situation we assume slow roll
 evolution before and after the transition, and a continuous potential
 at the transition. Therefore $H$ is always slow roll evolving, and
 there is no transition in the tensor modes power. For this reason, we
 simply assume $P_T(k) = r\,A\, (k/k_0)^{n_t}$
 and constrain the tensor spectral index to be given by the slow-roll consistency
 relation $n_t=-r/8$. 
 
\begin{figure}[t]
\begin{center}
\includegraphics[width=9.cm,trim=0cm 0cm 0cm  0cm,clip]{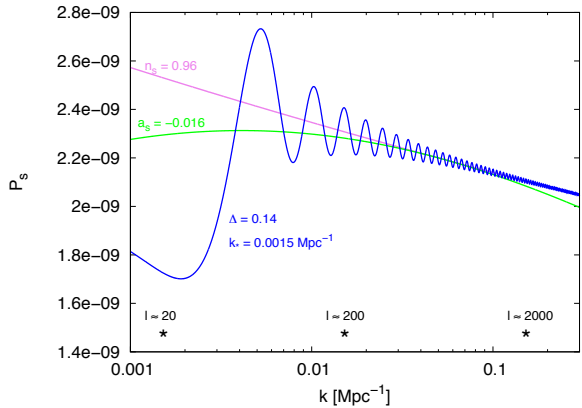}
\caption{Primordal scalar power spectrum with the best fit parameters for (i) a constant spectral index, (ii) a constant running, and (iii) the inflation  potential (\ref{starob}) with a  $\Delta = 0.14$ change in slope. This corresponds to a  $\sim 26\%$ suppression at large scales. The $\ell$ values indicated near the $x-$axis correspond to the multipole that is mostly sensitive to the mode with that momentum $k$. }
\label{fig:starob}
\end{center}
\end{figure}

\begin{figure}[t]
\begin{center}
\includegraphics[width=8.5cm,trim=0cm 0cm 0cm
  0cm,clip]{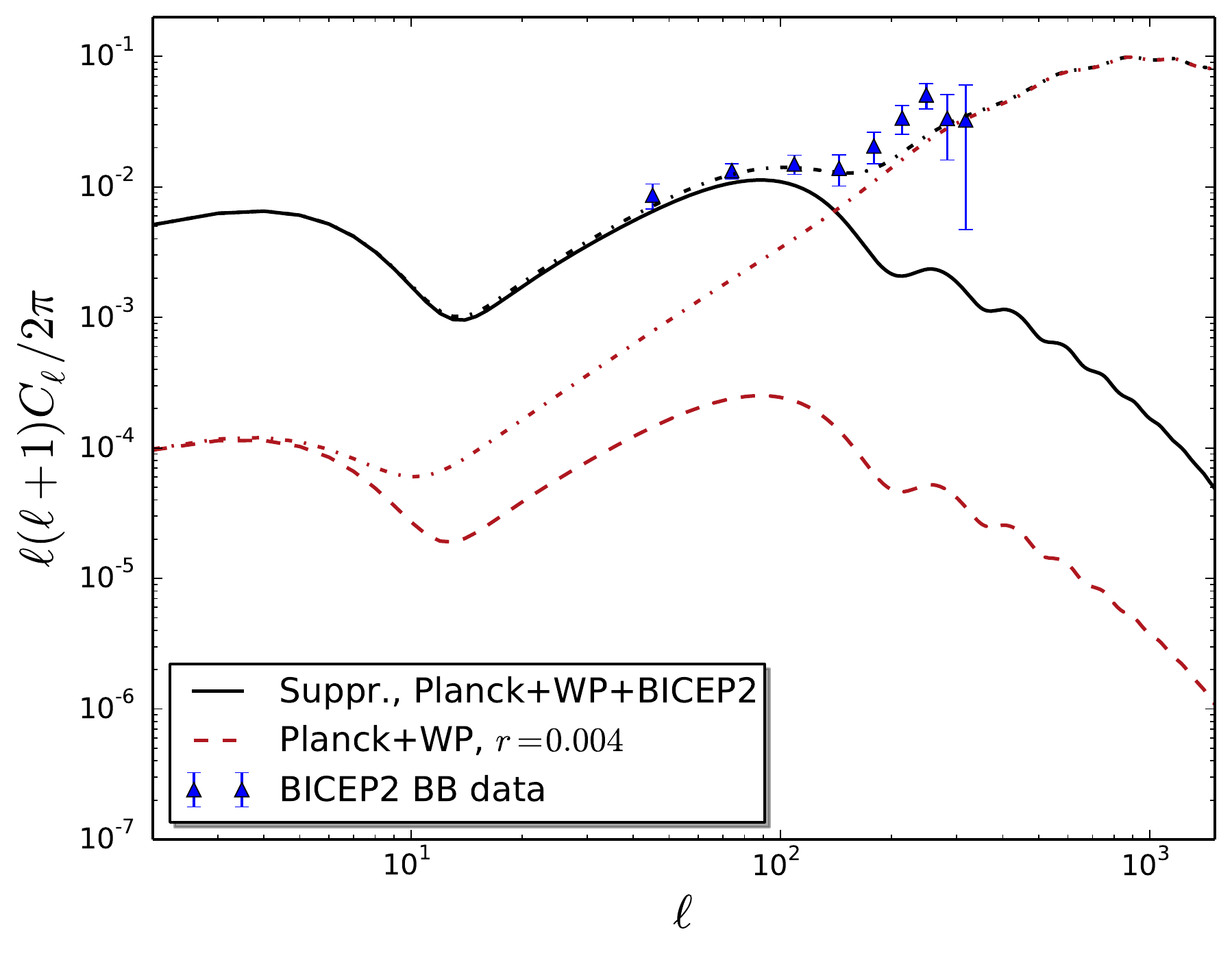}
  \caption{The $BB$ spectra for the same two models shown in
    Figure~\ref{fig:cltot}. The Planck ``base r'' best-fit model has
  a $B$-mode spectrum that is far below the BICEP2 detection, in the
un-lensed limit $\ell < 100$. The suppression model is compatible
with the detection having $r=0.18$ whilst giving a better fitting $TT$
spectrum. The best-fit parameters for the suppression model are
$\Delta=0.14$ and $k_\star=0.0015$ Mpc$^{-1}$.
}
  \label{fig:cltens}
   \end{center}
\end{figure}

Using a modified version of the {\tt CAMB} \cite{Lewis:1999bs} package we
calculate the total contribution to the $TT$ spectrum from both scalar
and tensor modes and the tensor contribution to the $BB$ spectrum. The
total contribution can then be fit to the combined Planck+WP+BICEP2
data set that include Planck $TT$ data, WMAP9 polarisation data, and
the new BICEP2 results. We do this by employing a modified version of
{\tt CosmoMC} \cite{Lewis:2002ah} with extra parameters $k_\star$ and
$\Delta$ included. For simplicity we carry out a MCMC exploration of
the reduced parameter space spanned by $\tau$ the optical depth to
reionization, $n_s$ the scalar spectral index, $\ln (10^{10} A)$ the
logarithm of the primordial curvature perturbation, $r$ the
tensor-to-scalar ratio, $k_\star$, and
$\Delta$ and keep all other cosmological and nuisance parameters fixed
to the best-fit values of the ``base r'' Planck+WP run reported in
\cite{planckXVI}. The ``base r'' $\Lambda$CDM + tensor model
will be the reference model for fit comparison.  We adopt uniform
prior ranges for the four parameters with $\tau=[0.01,0.8]$,
$n_s=[0.9,1.1]$, $\ln (10^{10}A) = [2.7,4]$, $r=[0,0.8]$, $k_\star=[0,0.015]$, and
$\Delta=[0,1.0]$. We also compare to the $\Lambda$CDM + tensor +
$\alpha_s$ model which has been highlighted as a way to reconcile the
high tensor amplitude with the $TT$ data. All runs include lensing
effects when calculating model $C_\ell$s.

\begin{figure}[t]
\begin{center}
\includegraphics[width=8.5cm,trim=0cm 0cm 0cm
  0cm,clip]{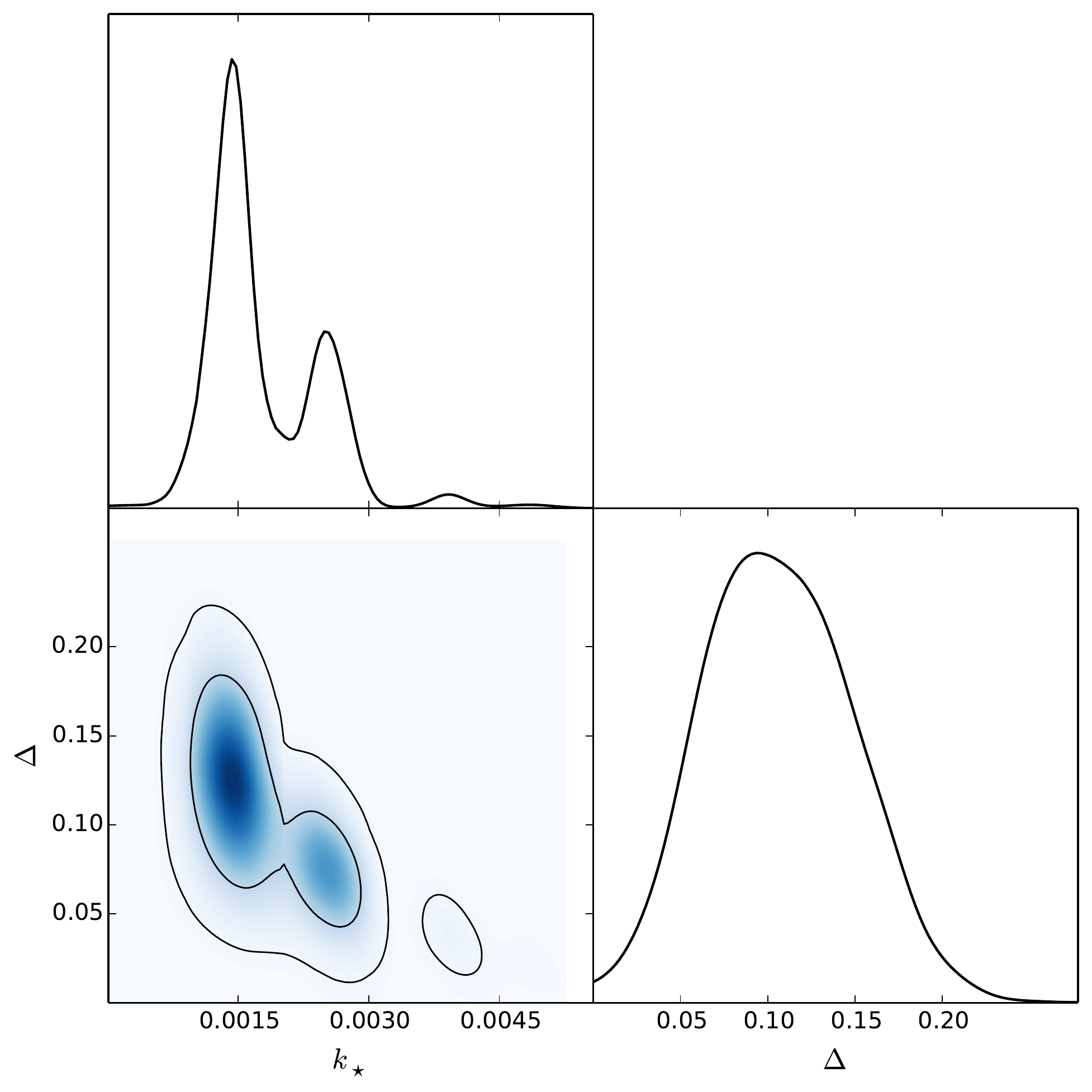}
  \caption{The marginalised, normalised, 1 and 2-D posteriors for $k_\star$ and
  $\Delta$. The parameters are well constrained by the $TT$ data. We recall that the best fit value $\Delta = 0.14$ corresponds to a $\sim 26\%$ suppression of the scalar  power at large scales. 
}
  \label{fig:marg}
   \end{center}
\end{figure}

\begin{table}[t]
\centering
\begin{tabular}{|l |c c c c|}
  \hline
  Planck+WP+BICEP2 & $\Delta N_p$ & $\chi^2$ & $\Delta \chi^2$& r \\[0.5ex]
  \hline
  $\Lambda$CDM + tensor & -- & 9853.104 & --& 0.18\\
  $\Lambda$CDM + tensor + $\alpha_s$ & +1 & 9846.426 & -6.68 & 0.18\\
  Suppression & +2 & 9838.87 & -14.23 & 0.18\\
  \hline
\end{tabular}
\caption{Relative changes in effective $\chi^2$ values for the three
  models considered. The change in number of model parameters with
  respect to the $\Lambda$CDM + tensor model is shown as $\Delta
  N_p$. The
  $r$ column shows the best-fit value of the tensor-to-scalar ratio
  for each case. Only the case with suppressed scalar power can
  accommodate the observed amplitude of $r\sim 0.2$.}
\label{tab:chi2}
\end{table}

Figure~\ref{fig:cltot} shows the total (scalar + tensor) contribution
to the $TT$ spectrum for the best-fit model obtain by Planck in their
``base r'' run \cite{planckXVI}, with $r=0.004$. This fit was obtained
with the Planck+WP combination only. We compare this with the best-fit
suppression model for the Planck+WP+BICEP2 data combination. The
best-fit suppression parameters are $k_\star=0.0015$ Mpc$^{-1}$,
$\Delta=0.14$, and the best-fit tensor-to-scalar ratio is
$r=0.18$. The suppression model has the freedom to obtain a better fit
than the conventional $\Lambda$CDM + tensor run by fitting the low
power observed at $\ell < 30$. In particular the model allows a better
fit to the low cluster of points in the range $20 < \ell <
30$. Including the BICEP2 results in the $\Lambda$CDM + tensor fit
forces the tensor amplitude up to a best-fit value of $r=0.18$. The
effective $\chi^2=-2\ln L$ for the best-fit is significantly higher
than the best-fit model in the suppression case. Table~\ref{tab:chi2}
shows the change in effective $\chi^2$ relative to the $\Lambda$CDM +
tensor model for fits including the Planck+WP+BICEP2 data
combination. The change in $\chi^2$ is indicative that the suppression
model is {\sl strongly} favoured over both standard and running
case.~\footnote{Concerning the running, we find the best fit value $\alpha_s = -0.016$ and the marginalized constraints $\alpha_s=-0.028 \pm 0.010$.} We attempt to quantify this by evaluating the Akaike Information
Criterion {\sl AIC} for the three models. The {\sl AIC} is defined as
\begin{equation}
  \mbox{\sl AIC} = 2\,N_p + \chi^2\,,
\end{equation}
where $N_p$ is the number of parameters in the model. It attempts to
properly take into account the penalty for using models with
increasing number of parameters to describe a set of data. The
relative likelihood between a model with minimum {\sl AIC} value and a
second model 
\begin{equation}
  L_{\mbox{\sl AIC}} = \exp\left[(\mbox{\sl AIC}_{\mbox{min}}-\mbox{\sl AIC})/2\right]\,,
\end{equation}
gives an estimate of the probability that the second model minimizes
the information required to describe the data\footnote{The {\sl
    corrected} {\sl AIC} also includes the effect of finite effective
  degrees of freedom $N_d$ in the data but we cannot use it here as it is very difficult
to define what $N_d$ is for the PLANCK+WP data combination since it is a
correlated data set whose likelihood is evaluated in both multipole
and pixel space. However we know that $N_d\sim{\cal O}(10^3)$ and the
correction goes as $\sim N_p^2/N_d$ and is therefore sub-dominant in
this case.}. The results of this analysis are summarised in
table~\ref{tab:AIC} and show the suppression model is strongly
favoured despite the extra parameters.

\begin{table}[t]
\centering
\begin{tabular}{|l |c c c|}
  \hline
  Planck+WP+BICEP2 & $N_p$ & $\Delta \mbox{\sl AIC}$& Rel. Likelihood\\[0.5ex]
  \hline
  $\Lambda$CDM + tensor & 21 &-10.617& $2.45\times 10^{-5}$\\
  $\Lambda$CDM + tensor + $\alpha_s$ & 22 & -5.939 & 0.0026\\
  Suppression & 23 & --  & 1.0\\
  \hline
\end{tabular}
\caption{Change in {\sl AIC} for the three models. The relative
  likelihood can be interpreted, in this case, as the probability that
the model minimizes the information required to describe the data
relative to the {\sl suppression} model that gives the best fit. The
{\sl AIC} attempts to properly penalise for any increase in number of
parameters in the models.}
\label{tab:AIC}
\end{table}

The results show that models with a suppression of scalar power at
large scales are favoured with respect to ones that employ running of
the scalar spectral index notwithstanding the additional parameter
required in the suppression model. Both models do better than the
$\Lambda$CDM + tensor model but the running case only marginally so
given the additional parameter. This is not surprising since the case with
running reduces power, for a negative $\alpha_s$, on scales that are
both lower {\sl and} higher than the pivot scale of $k=0.05$
Mpc$^{-1}$, see Figure \ref{fig:starob}. The model with running is therefore not a natural one to
appeal to if all that is required is to modify the scalar power on
large scales. 

Figure~\ref{fig:cltens} shows the $BB$ spectra for the same two models
as in Figure~\ref{fig:cltot}. The spectra include the pure tensor
contribution and the contribution from lensing. The BICEP2 data points
are included for comparison. The conventional ``base r'' Planck
best-fit model has a very low tensor spectrum which is incompatible
with the BICEP2 detection whilst the suppression model has a best-fit
value of $r=0.18$ in agreement with BICEP2.

The suppression parameters are well constrained by the Planck data, as
shown in Figure~\ref{fig:marg}, with marginalised constraints;
$k_\star=(1.80^{+0.86}_{-0.69})\times 10^{-3}$ Mpc$^{-1}$ and $\Delta
= 0.106^{+0.092}_{-0.013}$. The null hypothesis that the suppression is
either absent or occurs on scales larger than those observable seems
to be ruled out strongly.

\section{Scalar-tensor anticorrelations?}%
\label{anti}

Since the temperature fluctuations in the CMB take contributions from both scalar and tensor perturbations whereas the $B$-modes are sourced only by the latter, one can wonder whether it is possible to reduce the temperature power spectrum without affecting the $B$-mode one by introducing an (anti)correlation between primordial tensors and scalars. In this case one would not need to break primordial scale invariance to effectively suppress the temperature power spectrum {\em only at large scales} $\ell\lesssim 100$, since only at those scales the effect of tensors is not suppressed. On the other hand, such a correlation can be obtained only at the price of breaking Lorentz invariance. As a consequence, it would be interesting to explore possible connections to the large scale anomalies observed in WMAP and Planck. Here we will not study those implications but we focus on the main question of whether, and how, a primordial $\langle \zeta\,h_{ij}\rangle$ correlator can affect the temperature power spectrum. Nevertheless, we want to point out the following numerical coincidence. If the goal of the cross correlation is to reduce the impact of the tensor modes on the temperature anisotropies from $\sim 20\%$ to $\sim 10\%$, it should amount to $\sim 10\%$ of the total power. Interestingly, a $\la {\rm O } \left( 0.1 \right)$ amount of statistical anisotropy, only at the largest scales, appears to be present in the WMAP~\cite{Eriksen:2003db} and in the Planck~\cite{Ade:2013nlj} temperature data.

As we will show below, a primordial scalar-tensor correlation will affect the observable $C_{lm\,l'm'}\equiv \langle a_{lm}^*\,a_{l'm'}\rangle$. However, as noted in~\cite{Zibin:2014iea,Emami:2014xga} (see also~\cite{Chen:2014eua}), it will not affect the temperature power spectrum $C_l\equiv \frac{1}{2l+1}\sum_m \langle a_{lm}^*\,a_{lm}\rangle$, and will not, therefore, help suppressing the impact of a large value of $r$ on the temperature anisotropies as they are measured by Planck.

In this section we will study the effect on $C_{lm\,l'm'}$ of a simple form of scalar-tensor correlator, assuming that the primordial correlation functions are scale invariant and that the scalars affect the temperature fluctuations only through the Sachs-Wolfe effect.

We define the coefficients 
\begin{eqnarray}
&&C_{lm\,l'm'}\equiv \int\frac{d\bk\,d\bk'}{(2\pi)^3}\int d\Omega_\bp\,d\Omega_{\bp'}\,\langle \frac{\delta T}{T}(\bk,\,\bp)^*\,\frac{\delta T}{T}(\bk',\,\bp')\rangle\,Y_{l}^{m}(\bp)\,Y_{l'}^{m'}(\bp')^*\,,
\end{eqnarray}
where $\delta T$ will receive contributions both from the scalar perturbations and from the tensors: $\delta T=\delta T^s+\delta T^t$. In the equation above ${\bf p}$ is the unit vector directed along the line of sight.

The contribution to $C_{lm\,l'm'}$ from a primordial scalar-tensor correlation reads
\begin{eqnarray}\label{C-corr-par}
&&C^{\rm corr}_{lm\,l'm'}=\int\frac{d\bk\,d\bk'}{(2\pi)^3}\int d\Omega_\bp\,d\Omega_{\bp'}Y_{l}^{m}(\bp)\,Y_{l'}^{m'}(\bp')^*\nonumber\\
&&\times\left\langle\frac{\delta T^t}{T}(\bk,\,\bp)^*\frac{\delta T^s}{T}(\bk',\,\bp')+\frac{\delta T^s}{T}(\bk,\,\bp)^*\frac{\delta T^t}{T}(\bk',\,\bp')\right\rangle \;,
\end{eqnarray}%
that is proportional to the correlator between $\zeta$ and $h_{ij}$, that we assume to take the form
\begin{equation}\label{primordial-corr}
\langle\zeta(\bk)^* {h}_{ij}(\bk')\rangle=2\pi^2\,\frac{\delta(\bk-\bk')}{k^3}\,\Pi_{ij}{}^{ab}(\bk')\,\Theta_{ab} \;, 
\end{equation}
where $\Theta_{ab}$ is some given constant background tensor. For simplicity, we assume that  $\Theta_{ab}=\sigma\,{\bf v}_a\,{\bf v}_b$, for some constant vector ${\bf v}$ (the generalization to an arbitrary tensor is straightforward) and with $\sigma=\pm 1$. It is  then convenient to choose a coordinate system where ${\bf v}=(0,\,0,\,v)$. The term $\Pi_{ij}{}^{ab}$ is the projector on the transverse-traceless modes, as required by the tensor nature of $h_{ij}$: $\Pi_{ij}{}^{ab}=\Pi_i^a\,\Pi_j^b-\frac{1}{2}\Pi_{ij}\,\Pi^{ab}$ with $\Pi_{ab}=\delta_{ab}-\hat{k}_a\hat{k}_b$ and with $\hat\bk\equiv\bk/k$. 

An explicit calculation then shows that the only non vanishing coefficients are $C^{\rm corr}_{lm\,lm}$ and $C^{\rm corr}_{lm\,l+2m}$. The off-diagonal terms read
\begin{eqnarray}
&&C^{\rm corr}_{lm\,l+2m}=\frac{2\,\pi}{75}\,\frac{\sigma\,v^2}{l\,(l+1)}\,\frac{\left(7\,l^2+21\,l+27\right)}{(l+2)\,(l+3)\,(2\,l+3)}\,\sqrt{\frac{((l+1)^2-m^2)\,((l+2)^2-m^2)}{(2\, l+1) (2\, l+5)}} \;, \nonumber\\
\end{eqnarray}
whereas the diagonal terms are
\begin{eqnarray}
&&C^{\rm corr}_{lm\,lm}=-\frac{4\pi}{15}\,\frac{\sigma\,v^2}{l\,(l+1)\left(4\,l^2+4\,l-3\right)}\left(l^2+l-3\,m^2\right)\,.
\end{eqnarray}
Since $\frac{1}{2\,l+1}\sum_{m=-l}^l m^2=\frac{1}{3}\,l\,\left(l+1\right)$, the above result shows that the correlation has no effect on the coefficients $C_l\equiv\frac{1}{2l+1}\sum_m \langle a_{lm}^*\,a_{lm}\rangle$.

\section{Conclusions}
\label{conclusions}

The BICEP2 experiment has detected a $B$-mode polarization signal in
the CMB that can be explained by a lensed-$\Lambda$CDM + tensor
theoretical model, with tensor/scalar ratio
$r=0.2^{+0.07}_{-0.05}$. While keeping in mind that the central value
can be reduced with more statistics or with the subtraction of the
dust-polarized signal, we discussed some possible mechanism that may
explain how $r=0.2$ can be reconciled with the upper limits on tensor
modes from the temperature anisotropies measurements.

A possible way to reconcile this discrepancy is already offered in the
Planck \cite{Ade:2013uln} and the BICEP2 \cite{Ade:2014xna} analyses,
where a running of the spectral tilt is advocated. However, if the
central value $r=0.2$ will be confirmed, the required magnitude of the
running is significantly greater than the generic slow-roll
predictions. Motivated by this, we discussed two alternative
possibilities for suppressing the large scales temperature
anisotropies.

The first possibility involves a large scale suppression in the scalar
power. This can for instance be achieved if the inflaton zero mode has a greater
speed, or if the inflaton perturbations have a larger speed of sound, 
 when the largest observable scales are produced. This resembles
the study of \cite{Contaldi:2003zv}, in which the power at very large
drops to a negligible value due to a period of kinetic dominated
regime at the beginning  of the last $\sim 60 \,
e$-folds of inflation. Contrary to that case, here we study the
possibility that a partial drop of scalar power occurs at scales that
are larger than but comparable to the first acoustic peak. The best fit to the
temperature data is obtained for a transition scale $k_* \simeq 1.5
\times 10^{-3} \, {\rm Mpc}^{-1}$, with a $\simeq 26 \%$ decrease in
power. This fit improves over the Planck ones with conventional models
due to the suppression in power in the $10 \la \ell \la 30$
region. Moreover, the model easily fits the tensor-to-scalar ratio
$r\sim 0.2$ observed by BICEP2. Quite interestingly, the
$\chi^2$ of the fit of this model to the Planck+WP+BICEP2 data
improves by about $\sim 7.5$  with respect to the fit of the same data
of a model with constant running. This is a considerable improvement,
given that the suppression model has only one more parameter than the
model with a constant running. The significance of the improvement is supported by the Akaike Information Criterion, see Table \ref{tab:AIC}. 

A second possibility is a negative correlation between tensor and
scalar modes. Such a correlation is expected to occur, to some degree, in models with broken rotational symmetry. Such a situation actually appears to be present in the WMAP \cite{Eriksen:2003db}
and in the Planck \cite{Ade:2013nlj} temperature data. Unfortunately, a primordial tensor-scalar correlation does not affect the temperature power spectrum, that is used by Planck to set limits on the tensor-to-scalar ratio, and cannot therefore help reconcile the bounds on $r$ found from BICEP2 with those from Planck. It would be interesting to study whether, and under what conditions, such a conclusion might be avoided.\\

\appendix

\section{Step-function suppression of the scalar power} 
\label{app}

In this Appendix we consider the possibility of a step-function suppression of the scalar power: 
\begin{equation} 
P_s \left( k \right) = A \beta_s \, k^{n_s-1} \;, 
\label{parametrization-step}
\end{equation}
with $\beta_s = 1$ for $k \geq k_*$ and $\beta_s < 1$ for $k<k_*$. This parametrization should be compared for instance with the power spectrum (\ref{P-starob}) obtained in the Starobinsky model (\ref{starob}), with the identification $\beta_s \equiv \left(  1 - \Delta \right)^2$. As compared to that model, this parametrization does not have the ``ringing'' in the power visible in Figure \ref{fig:starob}. One could imagine alternative specific models that result in the suppression of power at large scales, as for instance a varying speed sound of the inflaton perturbations  \cite{Park:2012rh}. We expect that each different model will result in a different ``ringing'' structure. 
It is therefore instructive to compare the results presented in the main text with the results for a parametrization in which the ringing is absent.

\begin{figure}[t]
\begin{center}
\includegraphics[width=8.5cm,trim=0cm 0cm 0cm
  0cm,clip]{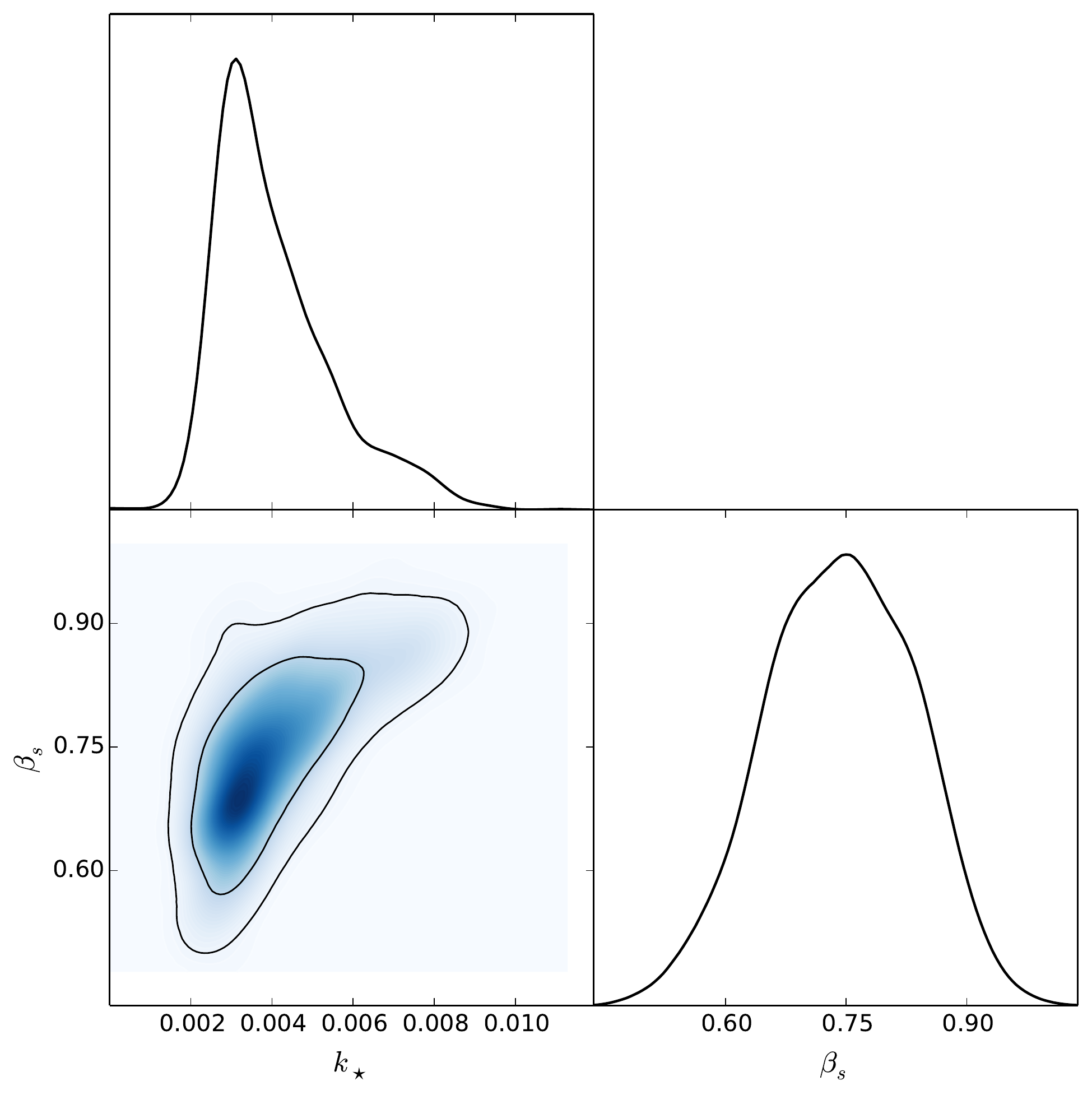}
 \caption{The marginalised, normalised, 1 and 2-D posteriors for $k_\star$ and  $\beta_s$ in eq. (\ref{parametrization-step}).
}
  \label{fig:marg-step}
   \end{center}
\end{figure}

We compare the parametrization (\ref{parametrization-step}) with the
data, and obtain the results summarized in Figure
\ref{fig:marg-step}. This result should be compared with the one shown
in Figure \ref{fig:marg} for the Starobinsky model. In this case we
obtain marginalised constraints $k_* = \left( 4.07_{-2.45}^{+0.70}
\right) \times 10^{-3} \, {\rm Mpc}^{-1} $ and $\beta_s =
0.74_{-0.09}^{+0.11}$, with an improvement $\Delta \chi^2 = - 14.23$
over the $\Lambda{\rm CDM} + {\rm tensor}$ fit. Upon the
identification $\beta_s \equiv \left( 1 - \Delta \right)^2$, the
result for $\beta_s$ translates into $\Delta = 0.14_{-0.06}^{+0.05}$.
Both values of $k_*$ and $\Delta$ are consistent with 
those obtained in the main text for the Starobinsky model
(\ref{starob}). Moreover, the improvement in $\chi^2$ is practically
identical to that obtained in the main text for the more physically
defined Starobinsky model. The small impact of the ringing effect on
the data analysis is due to the fact that a range of values of $k$
contributes to any given $C_\ell$, and this softens the impact of the
oscillation in the primordial power spectrum. This conclusion may not
apply for models that present more marked oscillations.

\begin{center}
{\bf Acknowledgements}\\
\end{center}

We thank Shaul Hanany, Clement Pryke, Gianmassimo Tasinato for very useful discussions. The work of  M.P. was partially supported  by DOE grant DE-FG02-94ER-40823 at the University of Minnesota. The work of L.S. was partially supported by the U.S. National Science Foundation grant PHY-1205986.


\begin{thebibliography}{99}


\bibitem{Ade:2014xna} 
  P.~A.~R. Ade {\it et al.}  [BICEP2 Collaboration],
  arXiv:1403.3985 [astro-ph.CO].

\bibitem{antibicep}
  H.~Liu, P.~Mertsch and S.~Sarkar,
  arXiv:1404.1899 [astro-ph.CO]; 
%
  M.~J.~Mortonson and U.~Seljak,
  arXiv:1405.5857 [astro-ph.CO].
%
  R.~Flauger, J.~C.~Hill and D.~N.~Spergel,
  arXiv:1405.7351 [astro-ph.CO].

\bibitem{Ade:2013uln} 
  P.~A.~R.~Ade {\it et al.}  [Planck Collaboration],
  arXiv:1303.5082 [astro-ph.CO].




\bibitem{Kosowsky:1995aa}
  A.~Kosowsky and M.~S.~Turner,
  ``CBR anisotropy and the running of the scalar spectral index,''
  Phys.\ Rev.\ D {\bf 52} (1995) 1739
  [astro-ph/9504071].



\bibitem{Chung:2003iu}
  D.~J.~H.~Chung, G.~Shiu and M.~Trodden,
  ``Running of the scalar spectral index from inflationary models,''
  Phys.\ Rev.\ D {\bf 68} (2003) 063501
  [astro-ph/0305193].
  



\bibitem{Easther:2006tv} 
  R.~Easther and H.~Peiris,
  ``Implications of a Running Spectral Index for Slow Roll Inflation,''
  JCAP {\bf 0609}, 010 (2006)
  [astro-ph/0604214].

\bibitem{Hannestad:2000tj} 
  S.~Hannestad, S.~H.~Hansen and F.~L.~Villante,
  ``Probing the power spectrum bend with recent CMB data,''
  Astropart.\ Phys.\  {\bf 16}, 137 (2001)
  [astro-ph/0012009].

\bibitem{Feng:2003mk} 
  B.~Feng, M.~-z.~Li, R.~-J.~Zhang and X.~-m.~Zhang,
  ``An inflation model with large variations in spectral index,''
  Phys.\ Rev.\ D {\bf 68}, 103511 (2003)
  [astro-ph/0302479].
  
\bibitem{BenDayan:2009kv} 
  I.~Ben-Dayan and R.~Brustein,
  ``Cosmic Microwave Background Observables of Small Field Models of Inflation,''
  JCAP {\bf 1009}, 007 (2010)
  [arXiv:0907.2384 [astro-ph.CO]].

\bibitem{Kobayashi:2010pz} 
  T.~Kobayashi and F.~Takahashi,
  ``Running Spectral Index from Inflation with Modulations,''
  JCAP {\bf 1101}, 026 (2011)
  [arXiv:1011.3988 [astro-ph.CO]].

\bibitem{Takahashi:2013tj} 
  F.~Takahashi,
  ``The Spectral Index and its Running in Axionic Curvaton,''
  JCAP {\bf 1306}, 013 (2013)
  [arXiv:1301.2834, arXiv:1301.2834 [astro-ph.CO]].

\bibitem{Kawasaki:2003zv} 
  M.~Kawasaki, M.~Yamaguchi and J.~'i.~Yokoyama,
  ``Inflation with a running spectral index in supergravity,''
  Phys.\ Rev.\ D {\bf 68}, 023508 (2003)
  [hep-ph/0304161].

\bibitem{Huang:2003zp} 
  Q.~-G.~Huang and M.~Li,
  ``CMB power spectrum from noncommutative space-time,''
  JHEP {\bf 0306}, 014 (2003)
  [hep-th/0304203].

\bibitem{BasteroGil:2003bv} 
  M.~Bastero-Gil, K.~Freese and L.~Mersini-Houghton,
  ``What can WMAP tell us about the very early universe? New physics as an explanation of suppressed large scale power and running spectral index,''
  Phys.\ Rev.\ D {\bf 68}, 123514 (2003)
  [hep-ph/0306289].

\bibitem{Yamaguchi:2003fp} 
  M.~Yamaguchi and J.~'i.~Yokoyama,
  ``Chaotic hybrid new inflation in supergravity with a running spectral index,''
  Phys.\ Rev.\ D {\bf 68}, 123520 (2003)
  [hep-ph/0307373].


\bibitem{Ballesteros:2005eg} 
  G.~Ballesteros, J.~A.~Casas and J.~R.~Espinosa,
  ``Running spectral index as a probe of physics at high scales,''
  JCAP {\bf 0603}, 001 (2006)
  [hep-ph/0601134].

\bibitem{Linde:1983gd} 
  A.~D.~Linde,
  Phys.\ Lett.\ B {\bf 129}, 177 (1983).

\bibitem{Freese:1990rb} 
  K.~Freese, J.~A.~Frieman and A.~V.~Olinto,
  Phys.\ Rev.\ Lett.\  {\bf 65}, 3233 (1990).


\bibitem{Kim:2004rp} 
  J.~E.~Kim, H.~P.~Nilles and M.~Peloso,
  JCAP {\bf 0501}, 005 (2005)
  [hep-ph/0409138].

\bibitem{Kaloper:2011jz} 
  N.~Kaloper, A.~Lawrence and L.~Sorbo,
  JCAP {\bf 1103}, 023 (2011)
  [arXiv:1101.0026 [hep-th]].

\bibitem{Kaloper:2008fb} 
  N.~Kaloper and L.~Sorbo,
  Phys.\ Rev.\ Lett.\  {\bf 102}, 121301 (2009)
  [arXiv:0811.1989 [hep-th]].


\bibitem{Pajer:2013fsa} 
  E.~Pajer and M.~Peloso,
  Class.\ Quant.\ Grav.\  {\bf 30}, 214002 (2013)
  [arXiv:1305.3557 [hep-th]].



\bibitem{Contaldi:2003zv} 
  C.~R.~Contaldi, M.~Peloso, L.~Kofman and A.~D.~Linde,
  JCAP {\bf 0307}, 002 (2003)
  [astro-ph/0303636].

\bibitem{Bennett:2003bz} 
  C.~L.~Bennett {\it et al.}  [WMAP Collaboration],
  Astrophys.\ J.\ Suppl.\  {\bf 148}, 1 (2003)
  [astro-ph/0302207].
  
  
\bibitem{Linde:2003hc} 
  A.~D.~Linde,
  JCAP {\bf 0305}, 002 (2003)
  [astro-ph/0303245].
  
  
  
\bibitem{Nicholson:2007by} 
  G.~Nicholson and C.~R.~Contaldi,
  JCAP {\bf 0801}, 002 (2008)
  [astro-ph/0701783].
  
\bibitem{Hazra:2013nca} 
  D.~K.~Hazra, A.~Shafieloo and G.~F.~Smoot,
  JCAP {\bf 1312}, 035 (2013)
  [arXiv:1310.3038 [astro-ph.CO]].


\bibitem{Starobinsky:1992ts} 
  A.~A.~Starobinsky,
  JETP Lett.\  {\bf 55}, 489 (1992)
  [Pisma Zh.\ Eksp.\ Teor.\ Fiz.\  {\bf 55}, 477 (1992)].

\bibitem{Kaloper:2003nv} 
  N.~Kaloper and M.~Kaplinghat,
  Phys.\ Rev.\ D {\bf 68}, 123522 (2003)
  [hep-th/0307016].


\bibitem{Park:2012rh} 
  M.~Park and L.~Sorbo,
  Phys.\ Rev.\ D {\bf 85}, 083520 (2012)
  [arXiv:1201.2903 [astro-ph.CO]].

\bibitem{Wu:2006xp} 
  C.~-H.~Wu, K.~-W.~Ng, W.~Lee, D.~-S.~Lee and Y.~-Y.~Charng,
  JCAP {\bf 0702}, 006 (2007)
  [astro-ph/0604292].
 
 
\bibitem{D'Amico:2013iaa} 
  G.~D'Amico, R.~Gobbetti, M.~Kleban and M.~Schillo,
  JCAP {\bf 1311}, 013 (2013)
  [arXiv:1306.6872 [astro-ph.CO]].
  
  
\bibitem{Zibin:2014iea} 
  J.~P.~Zibin,
  arXiv:1404.4866 [astro-ph.CO].

\bibitem{Emami:2014xga} 
  R.~Emami, H.~Firouzjahi and Y.~Wang,
  arXiv:1404.5112 [astro-ph.CO].

\bibitem{Chen:2014eua} 
  X.~Chen, R.~Emami, H.~Firouzjahi and Y.~Wang,
  arXiv:1404.4083 [astro-ph.CO].
  
  
\bibitem{Gumrukcuoglu:2006xj} 
  A.~E.~Gumrukcuoglu, C.~R.~Contaldi and M.~Peloso,
  astro-ph/0608405.
  

\bibitem{Pereira:2007yy} 
  T.~S.~Pereira, C.~Pitrou and J.~-P.~Uzan,
  JCAP {\bf 0709}, 006 (2007)
  [arXiv:0707.0736 [astro-ph]].


\bibitem{Gumrukcuoglu:2007bx} 
  A.~E.~Gumrukcuoglu, C.~R.~Contaldi and M.~Peloso,
  JCAP {\bf 0711}, 005 (2007)
  [arXiv:0707.4179 [astro-ph]].

\bibitem{Pitrou:2008gk} 
  C.~Pitrou, T.~S.~Pereira and J.~-P.~Uzan,
  JCAP {\bf 0804}, 004 (2008)
  [arXiv:0801.3596 [astro-ph]].

\bibitem{Martin:2011sn} 
  J.~Martin and L.~Sriramkumar,
  JCAP {\bf 1201}, 008 (2012)
  [arXiv:1109.5838 [astro-ph.CO]].

\bibitem{Lewis:1999bs} 
  A.~Lewis, A.~Challinor and A.~Lasenby,
  Astrophys.\ J.\  {\bf 538}, 473 (2000)
  [astro-ph/9911177].



\bibitem{Lewis:2002ah} 
  A.~Lewis and S.~Bridle,
  Phys.\ Rev.\ D {\bf 66}, 103511 (2002)
  [astro-ph/0205436].

\bibitem{planckXVI} Planck
Collaboration, Ade, P.~A.~R., Aghanim, N., et al.\ 2013,
arXiv:1303.5076

\bibitem{AIC} Akaike, H.\ 1974, IEEE 
Transactions on Automatic Control, 19, 716 
  
\bibitem{Eriksen:2003db} 
  H.~K.~Eriksen, F.~K.~Hansen, A.~J.~Banday, K.~M.~Gorski and P.~B.~Lilje,
  Astrophys.\ J.\  {\bf 605}, 14 (2004)
  [Erratum-ibid.\  {\bf 609}, 1198 (2004)]
  [astro-ph/0307507].


\bibitem{Ade:2013nlj} 
  P.~A.~R.~Ade {\it et al.}  [Planck Collaboration],
  arXiv:1303.5083 [astro-ph.CO].


  

  
  
\end{thebibliography}
\end{document}